# FCNHSMRA_HRS: Improve the performance of the movie hybrid recommender system using resource allocation approach

**Mostafa Khalaji[1] . Nilufar Mohammadnejad[2]**

1. Faculty of Computer Engineering, K .N. Toosi University of Technology, Tehran, Iran,
   Khalaji@email.kntu.ac.ir
2. Islamic Azad University, Shahr-e-Qods Branch, Tehran, Iran,
   Nilufar.mohd12@gmail.com

## Abstract

Recommender systems are systems that are capable of offering the most suitable services and products to users. Through specific methods and techniques, the recommender systems try to identify the most appropriate items, such as types of information and goods and propose the closest to the user's tastes. Collaborative filtering offering active user suggestions based on the rating of a set of users is one of the simplest and most comprehensible and successful models for finding people in the same tastes in the recommender systems. In this model, with increasing number of users and movie, the system is subject to scalability. On the other hand, it is important to improve the performance of the system when there is little information available on the ratings. In this paper, a movie hybrid recommender system based on FNHSM_HRS structure using resource allocation approach called FCNHSMRA_HRS is presented. The FNHSM_HRS structure was based on the heuristic similarity measure (NHSM), along with fuzzy clustering. Using the fuzzy clustering method in the proposed system improves the scalability problem and increases the accuracy of system suggestions. The proposed systems is based on collaborative filtering and, by using the heuristic similarity measure and applying the resource allocation approach, improves the performance, accuracy and precision of the system. The experimental results using MAE, Accuracy, Precision and Recall metrics based on MovieLens dataset show that the performance of the system is improved and the accuracy of recommendations in comparison of FNHSM_HRS and collaborative filtering methods that use other similarity measures for finding similarity, is increased.

## 1. Introduction

By increasing information on the Internet and cyberspace and online purchases and user interactions, recommender systems(RS) have been studied to guide users towards their desires or needs over the last 20 years, especially in the first decade of the twenty-first century, and research many have done this [1]. Recommender systems are categorized according to how they are offered to different models, one of the most important of which is the collaborative filtering. The performance of this method is to provide an active user based on a similarity between users or items. Hence, it involves memory-based and model-based techniques [2]. The memory-based model initially calculates the similarity between the users and then selects the same users as the active user as neighboring users and finally offers the active user based on these users. The model-based collaborative filtering initially creates a model of user behavior and then predicts unrated movies based on that model. Cold start and scalability are problematic in the recommender systems. When the recommender system encounters a lack of information (ratings) from the user's history, the cold start problem is for new user in the system, and if a new movie is logged in, since users have not already seen the movie before, the cold start problem for the new movie will happen. On the other hand, with the dramatic increase in the number of users in cyberspace, the scalability problem for recommender systems occurs, which means that the system will lose its performance and accuracy to some extent. Recommender systems that use collaborative filtering, are very easy and simple to understand and to implement.

In this paper, the main focus of the proposed system on system performance is using a combination of memory-based and model-based methods to solve the problem of scalability. The core of this system is based on the structure of the FNHSM_HRS hybrid recommender system, which initially uses fuzzy clustering and collaborative filtering algorithms [3], which is based on the NHSM[1] measure introduced in reference [4]. To determine the users of the neighbor with the active user, and then by adding a resource allocation, which is one of the basic issues in social networks, the degree of reliability is determined by the similarity of users with the active user.

The use of clustering techniques in this type of recommender system makes users who are similar in tastes to each other cluster in specific groups, and then the prediction and suggestion operations of movies in each cluster separately. The cold start problem is due to the lack of sufficient information(ratings) on the system by the user or the movie. The resource allocation task is that no matter how much the two $x$ and $y$ users observe or select popular movies, the reliability degree between them is less and the more unlikely they are interested in movies, the more reliability degree they will be more reliable [5]. Therefore, this improves the performance of the recommender system in terms of time and accuracy.

The structure of the article is as follows. In the second part, we will present the summary of recent research done by researchers. The third part introduces the proposed recommender system. The fourth part is the evaluation section and the implementation results, in which the results of tests and comparison with other methods are dealt with. Finally, a conclusion is made in the final section.

## 2. Related Works

Recommender systems were initially introduced by Goldberg et.al. [6] Subsequently, several methods were proposed based on the proposed method, one of which is the collaborative filtering. Collaborative filtering is one of the most popular privatization offerings for users that is used in many areas. This technique performs its operations based on a series of similarity and series of predefined models. Although this method suffers from problems such as cold start, data sparsity and scalability, it is very easy to understand and implement them, and they are the base models in the recommender systems.

---

[1] New Heuristic Similarity Measure

To improve system performance, many researchers have introduced a variety of similarity measures. These measures include Pearson, Cosine, PIP and so on [7, 8].

Bellogin introduced methods to improve the performance of the recommender systems, which selected Herlocker's weighting and McLaughlin's weighting methods to determine which users were closely related to the user's tastes [9]. Haifeng Et.al also proposed a new heuristic similarity measure, which considered three aspects of proximity, impact and popularity of user ratings when choosing the neighboring users for the active user. On the other hand, a combined measure of Jaccard and Mean Squared Difference was presented by Bobadila et al. [10].

Using clustering techniques to group similar users in terms of tastes helps to solve scalable issues. Koohi and his colleague, using fuzzy clustering and maximization method, for example, by assigning users to clusters with different membership degrees and using Pearson similarity measure to find the closest neighbors, showed that their performance of system against K-Means and SOM techniques have been improved [11]. Khalaji and his colleagues proposed a hybrid recommender system (FNHSM_HRS)[2] that was a combination of fuzzy clustering technique and heuristic similarity measure, which improved the performance of the recommender system compared to traditional systems [3].

The resource allocation method was introduced by Liben-Nowell and Kleinberg in 2007 [12]. Resource allocation (RA) is a well-known method in link prediction problems where the likelihood of existence of a link between non-connected nodes is estimated. There are a number of methods for link prediction in complex networks [13]. RA method is used to predict missing links or possible future link. Many issues in social networks and data mining can be modeled on a problem of predicting communications between users. For example, friend suggestion on social networks [14], the issue of product offerings on online systems [15], and so on. In 2017, Khalaji designed a hybrid recommender systems based on neural network and resource allocation that solved the problem of scalability and cold start [5].

## 3. FCNHSMRA_HRS

Figure 1 shows the structure of the proposed system FCNHSMRA_HRS, which is derived from the FNHSM_HRS structure along with the resource allocation approach. In this section, we discuss about the propose method which FCNHSMRA_HRS (Fuzzy Clustering, New Heuristic Similarity Measure, Resource Allocation, Hybrid Recommender System) consists of two phases: offline and online. In the offline phase, the preprocessing on the data is done, for example, by training system model based on information the user-movie ratings matrix, which is the same as the FNHSM_HRS structure. Then the prediction based on NHSM and RA approaches is made on the online phase.

### 1-3- Offline Phase

The traditional methods of recommender systems only help to find users who are in common tastes with active user in a variety of similarity measure, which is not effective when the user ratings matrix is sparse. On the other hand, most of these measures suffer from the complexity of time. To solve these problems, by clustering users in different clusters in the proposed system, it accelerates the process of system operation and suggests that users in each cluster are similar in tastes to one another.

---

[2] Fuzzy NSHM_Hybrid RS

In this section, the ratings matrix is given as input to the recommender system; 80% of the ratings matrix is selected as the training dataset. Using the fuzzy clustering method, system users are divided into a number of clusters. In the proposed system, according to the researchers' suggestions [11], 3 clusters are considered. Fuzzy clustering algorithm by assigning a different membership degree to users in each cluster specifies the impact of users on each cluster. For ranking users and selecting users in the system's prediction phase, we use one of the defuzzification methods called COG[3].

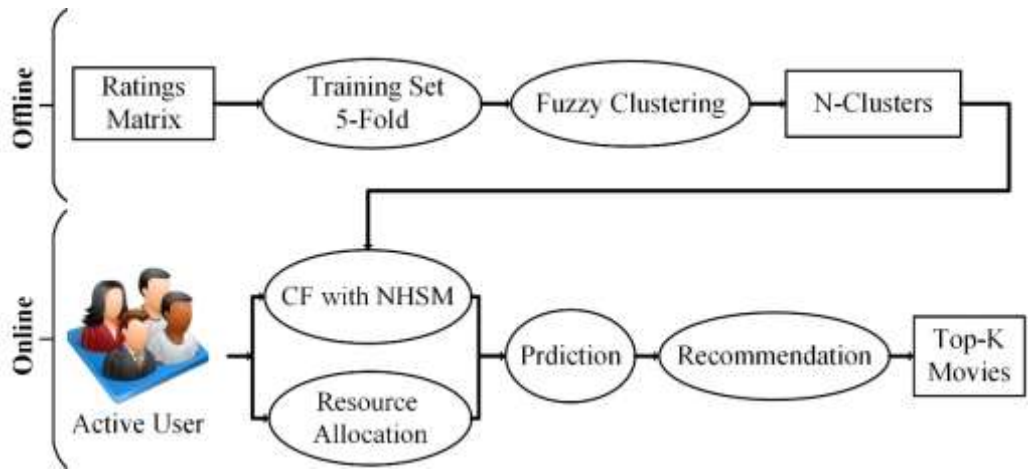

**Figure 1 : The FCNHSMRA_HRS Structure**

### 1-3- Online Phase

In this section, the active user who is scheduled to submit suggestions on unobserved movies is logged and, based on his active user history, calculates the closest neighbor in the active user cluster by the NHSM. This measure has two main coefficients that are specified in equation (1).

$$NHSM\_Sim(u,v) = JPSS\_Sim(u,v) \cdot URP_{Sim(u,v)} \quad (1)$$

To calculate the $NHSM\_Sim$ similarity measure, the $JPSS\_Sim$ similarity measure should first be calculated, which itself is derived from two similarity measures that are mentioned in equations (2) and (3).

$$JPSS_{Sim(u,v)} = PSS_{Sim(u,v)} \cdot Jaccard'_{Sim(u,v)} \quad (2)$$

$$Jaccard'_{Sim(u,v)} = \frac{|I_u \cap I_v|}{|I_u| \times |I_v|} \quad (3)$$

---

[3] Center of gravity

$|I_u \cap I_v|$ represents the number of similar movies that users $u$ and $v$ have seen and $|I_u|$ indicates the number of movies that the active user ($u$) and $|I_v|$ indicates the number of movies the other user ($v$) have rated. The $PSS\_Sim$ measure is obtained through equation (4).

$$PSS_{Sim}(r_{u,p}, r_{v,p}) = Proximity(r_{u,p}, r_{v,p}) . Significance(r_{u,p}, r_{v,p}) . Singularity(r_{u,p}, r_{v,p}) \quad (4)$$

$Proximity$, $Significance$ and $Singularity$ are calculated through equation (5-7).

$$Proximity(r_{u,p}, r_{v,p}) = 1 - \frac{1}{1 + \exp(-|r_{u,p} - r_{v,p}|)} \quad (5)$$

$$Significance(r_{u,p}, r_{v,p}) \frac{1}{1 + \exp(-|r_{u,p} - r_{med}| . |r_{v,p} - r_{med}|)} \quad (6)$$

$$Singularity(r_{u,p}, r_{v,p}) = 1 - \frac{1}{1 + \exp(-|\frac{r_{u,p} + r_{v,p}}{2} - \mu_p|)} \quad (7)$$

Where $r_{u,p}$ is the rating of movie $p$ by user $u$ and $r_{v,p}$ is the rating of movie $p$ by user $v$. $r_{med}$ is the median value in the rating scale. The rating matrix in the FCNHSMRA_HRS recommender system has a scale of 1-5, with an average of 3. $\mu_p$ is also the average rating of movie $p$ by users.

The last step in equation (1) is to calculate the similarity measure of $URP\_Sim$, which is obtained by using equation (8).

$$URP_{Sim(r_{u,p}, r_{v,p})} = 1 - \frac{1}{1 + \exp(-|\mu_u - \mu_v| . |\sigma_u - \sigma_v|)} \quad (8)$$

Where $\mu_u$ is the mean of rating of user $u$ and $v$ respectively. $\sigma_u$ and $\sigma_v$ represent the standard variance of user $u$ and $v$, which is obtained in accordance with equation (9).

$$\sigma_u = \sqrt{\sum_{p \in I_u}(r_{u,p} - \overline{r_u})^2 / |I_u|} \quad (9)$$

After calculating the similarity measure, the operation of calculating the reliability degree between the user and the active user is calculated, this degree determines how much can be trusted in the degree of similarity obtained from the NHSM between the active user and the neighboring user for the closeness of their tastes [5]. Therefore, considering equation (10), the operation of determining the reliability degree between the two users is calculated.

$$R_{RA}(u_i, u_j) = \sum_{z \in \Gamma(u_i) \cap \Gamma(u_j)} \frac{1}{k_z} \quad (10)$$

Where $z$ is the observed movie between $u_i$ and $u_j$ and $\Gamma(u_i)$, the number of movies observed by user $i$, as well as $\Gamma(u_j)$, the number of movies observed by user $j$. $k_z$ is the degree of $z$ interconnection between the user $i$ and the user $j$, in other words $k_z$ is the $z$ movies grade, for example, the number of users who have rated the movie $z$.

Next step, the number of neighboring users will be specified and selected based on the highest degree of similarity of neighboring users to the active user in the corresponding cluster.

Therefore, according to equation (11), the prediction operations of the unrated movies are calculated based on collaborative filtering in the FCNHSMRA_HRS system.

$$Predict(u,i) = \mu_u + \frac{\sum_{j=1}^{m}(r_{v_j,i} - \mu_v) \cdot NHSM\_Sim(u,v_j) \cdot R_{RA}(u,v_j)}{\sum_{j=1}^{m}|NHSM_{Sim(u,v_j)} \cdot R_{RA}(u,v_j)|} \quad (11)$$

Where $u$ is the active user and $i$ is a movie that the FCNHSMRA_HRS system is supposed to predict a rating for that. The FCNHSMRA_HRS system calculates these ratings for all unobserved movies by the active user and offers the top $n$ movies. In equation (11), $\mu_u$ is the average ratings of active user and $m$ is the number of neighboring users within the active user cluster. $NHSM\_Sim(u,v_j)$ is the similarity degree of the active user $u$ with the user $v_j$. $r_{v,i}$ is the rating of movie $i$ by user $v$ and $\mu_v$ is the average ratings of user $v$.

## 4. Evaluation of FCNHSMRA_HRS system

The performance of the proposed system was evaluated in the MovieLens dataset which consists of 943 users and 1682 movies with 100,000 ratings for movies [16]. The ratings range in this dataset is from 1 to 5, which 5 being excellent and 1 being terrible.

To evaluate the system's performance, we tested on the dataset and used the 5-fold cross validation algorithm, which provides 80% of the data for training and creating a proposed system model and 20% for system testing. The system evaluation based on the MAE, Accuracy, Precision and Recall metrics has been calculated according to (12-15) on the test data, which are shown in Table 1 of the related confusion matrix [17].

| Actual / Predicted | Negative | Positive |
|---|---|---|
| Negative | A | B |
| Positive | C | D |

**Table 1: Confusion Matrix [17]**

$$MAE = \frac{\sum_{i=1}^{n}|\hat{r}_{u,i} - r_{u,i}|}{n} \quad (12)$$

$$Accuracy = \frac{Correct\ Recommendation}{Total\ Possible\ Recommedation} = \frac{A+D}{A+B+C+D} \quad (13)$$

$$Precision = \frac{Correctly\ Recommended\ Items}{Total\ Recommeded\ Items} = \frac{D}{B+D} \quad (14)$$

$$Recall = \frac{Correctly\ Recommended\ Items}{Total\ Useful\ Recommedations} = \frac{D}{C+D} \quad (15)$$

As we have already mentioned in the proposed system (FCNHSMRA_HRS), according to the FNHSM_HRS structure, the number of clusters was 3 and the number of neighboring users was 50. Each fold is repeated 5 times with independent running for each Top-N and the results are saved. In this system, a heuristic similarity measure has been used along with a resource allocation approach. As a result, five methods have been used such as: FNHSM_HRS, Pearson, Cosine, McLaughlin's weighting and the Herlocker's weighting methods [9]. You see the results in the Tables (2) and (3).

**Table 2: System Evaluation Results**

| Number | Method Name | Evaluation Metric | Top 5 Movies | Top 10 Movies | Top 15 Movies | Top 20 Movies | Top 30 Movies |
|---|---|---|---|---|---|---|---|
| 1 | **FCNHSMRA_HRS** | Accuracy | **64.146** | **61.510** | **59.317** | **56.002** | **55.814** |
|  |  | Precision | **93.468** | **92.010** | **91.888** | **90.946** | **90.891** |
|  |  | Recall | **60.701** | **55.115** | **51.642** | **48.090** | **47.327** |
|  |  | MAE | **0.741** | **0.759** | **0.771** | **0.779** | **0.783** |
| 2 | FNHSM_HRS [3] | Accuracy | 60.203 | 59.373 | 56.200 | 55.557 | 55.155 |
|  |  | Precision | 91.391 | 91.148 | 90.962 | 90.902 | 90.844 |
|  |  | Recall | 58.643 | 51.029 | 47.560 | 45.486 | 44.379 |
|  |  | MAE | 0.756 | 0.766 | 0.778 | 0.782 | 0.789 |
| 3 | F_CF with Pearson [3] | Accuracy | 59.537 | 56.756 | 55.784 | 55.160 | 54.762 |
|  |  | Precision | 90.674 | 90.352 | 90.230 | 90.147 | 90.082 |
|  |  | Recall | 54.537 | 50.090 | 46.790 | 44.834 | 42.768 |
|  |  | MAE | 0.768 | 0.775 | 0.782 | 0.787 | 0.794 |
| 4 | F_CF with Cosine [3] | Accuracy | 60.010 | 57.275 | 56.035 | 55.073 | 53.610 |
|  |  | Precision | 82.020 | 81.371 | 80.935 | 80.683 | 80.441 |
|  |  | Recall | 50.278 | 50.206 | 47.041 | 45.421 | 44.003 |
|  |  | MAE | 0.827 | 0.828 | 0.829 | 0.834 | 0.839 |
| 5 | F_MW [3] | Accuracy | 59.507 | 57.717 | 56.113 | 55.734 | 55.056 |
|  |  | Precision | 91.098 | 90.856 | 90.691 | 90.617 | 90.550 |
|  |  | Recall | 58.039 | 51.015 | 47.576 | 45.433 | 43.383 |
|  |  | MAE | 0.758 | 0.768 | 0.778 | 0.783 | 0.791 |
| 6 | F_HW [3] | Accuracy | 58.172 | 57.002 | 55.929 | 55.406 | 54.904 |
|  |  | Precision | 90.881 | 90.562 | 90.500 | 90.429 | 90.346 |
|  |  | Recall | 58.220 | 50.604 | 47.074 | 45.136 | 43.042 |
|  |  | MAE | 0.777 | 0.785 | 0.791 | 0.796 | 0.803 |

**Table 3: Average Top-N for each method**

| Number | Method Name | Evaluation Metric | Average of Top-N |
|---|---|---|---|
| 1 | **FCNHSMRA_HRS** | Accuracy | **59.3578** |
| | | Precision | **91.8406** |
| | | Recall | **52.575** |
| | | MAE | **0.7666** |
| 2 | FNHSM_HRS [3] | Accuracy | 57.2976 |
| | | Precision | 91.0494 |
| | | Recall | 49.4194 |
| | | MAE | 0.7742 |
| 3 | F_CF with Pearson [3] | Accuracy | 56.3998 |
| | | Precision | 90.297 |
| | | Recall | 47.8038 |
| | | MAE | 0.7812 |
| 4 | F_CF with Cosine [3] | Accuracy | 56.4006 |
| | | Precision | 81.09 |
| | | Recall | 47.3898 |
| | | MAE | 0.8312 |
| 5 | F_MW [3] | Accuracy | 56.8254 |
| | | Precision | 90.7624 |
| | | Recall | 49.0892 |
| | | MAE | 0.7756 |
| 6 | F_HW [3] | Accuracy | 56.2826 |
| | | Precision | 90.5472 |
| | | Recall | 48.8152 |
| | | MAE | 0.7904 |

## 5. Conclusion

The main purpose of the recommender systems is to offer a series of suggestions based on the user's tastes and to find users who are similar in tastes to the active user. Therefore, one of the main challenges of these systems is the accuracy and performance of these systems, given the large amount of information (increasing the number of users and system items) in the shortest possible time. Our aim in this paper is to improve the performance of the FNHSM_HRS hybrid recommender system. The proposed method in the proposed FCNHSMRA_HRS system, which consists of combining fuzzy clustering technique, the NHSM similarity measure, and applying the resource allocation approach, has been instrumental in improving Accuracy, Precision, Recall, and MAE in movie recommender systems. The results of the experiments on the proposed system FCNHSMRA_HRS with an average of MAE of **0.7666**, an Accuracy of **59.3578**, a Precision of **91.8406** and a Recall of **52.575**, which, in contrast to the FNHSM_HRS system and other similarity measures, indicate improved system performance and increased accuracy is. The proposed system of FCNHSMRA_HRS solved the problem of scalability in the recommender systems using the clustering method.